\documentclass[numreferences]{article}
\usepackage{latexsym,amsmath,amsthm,amssymb}

\usepackage{bm}

\usepackage{graphicx,epsfig}

\begin{document}

\title{Solving the brachistochrone and other variational problems with soap films.
}
\date{}
\maketitle

C. Criado{$\footnotesize{^{a)}}$}

{\small{\emph{Departamento de Fisica Aplicada I, Universidad de
Malaga, 29071 Malaga, Spain}}} \vskip.5cm N.
Alamo{$\footnotesize{^{b)}}$}

{\small{\emph{Departamento de Algebra, Geometria y Topologia,
Universidad de Malaga, 29071 Malaga, Spain}}}

\begin{abstract}
We show a method to solve the problem of the
brachistochrone as well as other variational problems with the help of the soap films that are formed between two suitable surfaces.
We also show the interesting connection between some variational problems of dynamics, statics, optics, and elasticity.
\end{abstract}

\section{Introduction}

The calculus of variations can be used in the formulation of most
of the physical problems. Its origin was the famous problem of the
\emph{brachistochrone}, the curve of shortest descent
time.{\footnotemark[1]} Johann Bernoulli found that the
curve solution to this problem is a cycloid, that is, the curve
described by a point on the circumference of a circle that rolls
along a straight line.{\footnotemark[2]}

The calculus of variations was also present in the solution of
other classical problems, such as the catenary and the
isoperimetric problem. Its development was
parallel to that of mechanics, geometrical optics, and elasticity.
In this article we show how these problems can be related with the
Plateau problem, that is, the problem of finding
the form that a soap film adopts for certain contour restrictions.{\footnotemark[3]}{$\footnotesize{^- }$}{\footnotemark[8]}
 We also
show how to design simple experiments with soap films to find the
solution to other variational problems such as a generalized Steiner
problem or the problem of the quickest approach.

This article was inspired by reading the book ``Demonstrating Science with Soap
Films" by D. Lovett, in which a similar argument is used to
generate the catenary.{\footnotemark[8]}
The paper can be useful for students and teachers of Physics
undergraduate courses. A complete study of this subject requires
the use of the  Euler-Lagrange equations. However, given the intuitive
character of the analogies between mechanics, optics,
and elasticity problems that we establish, together with the easy visualization of
the solutions with soap films, this article may also be useful for
students without knowledge of differential equations. In the Appendix we propose a problem to use the concepts that appear in the article in a new situation.

\section{\label{our model} Relation between the Brachistochrone and Plateau's problems.}

\begin{figure}
\begin{center}
\includegraphics[width=16pc]{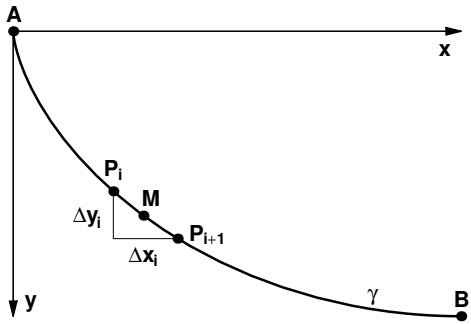}
\caption{The velocity of a heavy body after descending an altitude $y$ along $\gamma$ is  $\sqrt{2gy}$.}
\end{center} \label{Figure1}
\end{figure}

Johann Bernoulli posed the problem of the brachistochrone as follows (see Figure~1):

\emph{Let two points $A$ and $B$ be given in a vertical plane. To find the curve that a point $M$, moving on a path $AMB$, must follow that, starting from $A$, it reaches $B$ in the shortest time under its own gravity.}{\footnotemark[1]}

Denote by $\gamma$ a curve joining $A$ and $B$, and let us approximate this curve by a polygonal
$A=P_1 < \cdots < P_i <P_{i+1}< \cdots <P_n=B$,
such that the distance $\vert P_iP_{i+1}\vert$ is very small, thus
we can assume that the velocity along $P_iP_{i+1}$ is constant and
can be taken as the velocity of a heavy body after descending an
altitude $\bar{y}_i$, i.e. $v_i=\sqrt{2 g \bar{y}_i}$, where $g$
is the gravity acceleration and $\bar{y}_i$ is the ordinate of the midpoint between
 $P_i$ and $P_{i+1}$ (see Figure~1).

Then the total time of descent along the polygonal is given by:
\begin{equation}\label{}
    \sum_{i=1}^n \frac{\vert P_iP_{i+1}\vert}{v_i}=
    \sum_{i=1}^n \frac{\sqrt{\Delta x_i^2 + \Delta y_i^2}}{v_i}=
    \sum_{i=1}^n \frac{\sqrt{1 + (\frac{\Delta y_i}{\Delta x_i})^2}\Delta x_i}{\sqrt{2g\bar{y}_i}}
\end{equation}
The  total time of descent along $\gamma$, $t(\gamma)$,  is then
obtained from this approximation, taking the limit as the length of the
 segments of the polygonal goes to $0$, thus we obtain
\begin{equation}\label{Brachis}
    t(\gamma)= \frac{1}{\sqrt{2g}}\int_{\gamma}\frac{\sqrt{1+y'^2}}{\sqrt{y}}dx.
\end{equation}

The problem is to find the curve $\gamma$ that minimizes the above
expression. The solution to this
problem is given by solving a differential equation, known as
Euler-Lagrange equation, whose
solution is a cycloid.{\footnotemark[9]}

We are going to study a Plateau problem that, from a mathematical point
of view, is equivalent to the Brachistochrone problem and, therefore, its solution is also the cycloid.

\begin{figure}
\begin{center}
\includegraphics[width=26pc]{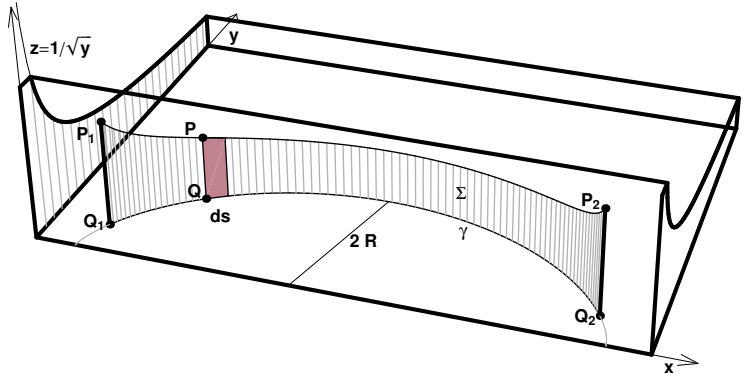}
\caption{Vertical surface $\Sigma$ that aproximates the soap film bounded by the surfaces $z=0$, $z=\frac{1}{\sqrt{y}}$, and two vertical segments. Its orthogonal projection is the cycloid $\gamma$.}
\end{center} \label{Figure2}
\end{figure}

Consider two surfaces, one is the horizontal plane $z=0$, and the other is defined by
the equation $z=\frac{1}{\sqrt{y}}$ (see Figure~2).
Consider also two vertical segments $P_1Q_1$ and $P_2Q_2$ connecting
the two surfaces.
We propose the problem of finding the surface of minimal area, bounded
by these two segments and the two surfaces.
A surface with such characteristics is known in mathematics as a minimal
 surface.{\footnotemark[10]} From the point of view of Physics,
 the surface corresponds to a soap film. This follows from the fact that
 the soap film must minimize the elastic energy, what is equivalent to
 minimize the area of the surface.

We will approximate this soap film by a vertical surface $\Sigma$,
that is, a surface that is projected orthogonally on a curve
$\gamma$ connecting points $Q_1$ and $Q_2$ in the plane $Oxy$.
As we shall see, this approximation fits very well the experimentally observed soap film.{\footnotemark[11]}

The area of the surface $\Sigma$ for a given curve $\gamma$ is:

\begin{equation}\label{Plateau}
    A(\gamma)=  \int_{\Sigma} dA =
    \int_{\gamma}\frac{1}{\sqrt{y}}ds=
     \int_{\gamma}\frac{\sqrt{1+y'^2}}{\sqrt{y}}dx,
\end{equation}
where the element of area $dA$ is the product of the length
element $ds = \sqrt{1+y'^2} dx$ of the curve $\gamma$ by
$\frac{1}{\sqrt{y}}$ (see Figure~2).

We observe that Eqs.~(\ref{Brachis}) and~(\ref{Plateau}) are identical except for a constant factor. This means
that the curve that minimizes the area $A(\gamma)$,
 is the same curve that minimizes the time of descent
$t(\gamma)$. Therefore, the solution curve is the same as for the
Brachistochrone problem, that is, a cycloid.

Figure~2 shows a representation of the surface $\Sigma$ and the
cycloid $\gamma$. The parametric equations of the cycloid in the
coordinates of the figure are:
\begin{equation}\label{}
    x=x_0 + R(\alpha - \sin \alpha)\ ;\quad y=R(1-\cos \alpha)\ :
    \quad 0\le \alpha \le 2\pi,
\end{equation}
where $x_0$ and $R$ can be determined by the coordinates of the
points $Q_1$ and $Q_2$.

We observe from Eq.~(\ref{Plateau}) that $A(\gamma)$ is also the
length of the curve $\gamma$ with the metric given by $ds^* =
\frac{1}{\sqrt{y}} ds$, that is, the Euclidean metric $ds=
\sqrt{dx^2 + dy^2}$ divided by $\sqrt{y}$. The above discussion
shows that the cycloid is the curve along which the distance
between two points $Q_1$, $Q_2$ for the metric $ds^*$ is
minimized. For this reason, $ds^*$ is known as the cycloidal
metric.

\begin{figure}
\begin{center}
\includegraphics[width=26pc]{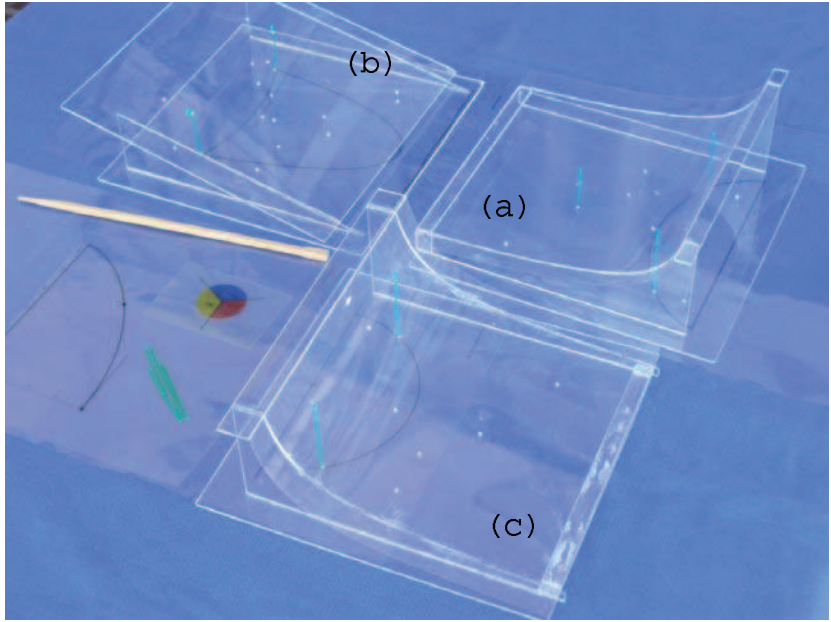}
\caption{Perspex models.}
\end{center} \label{Figure3}
\end{figure}

The perspex model (a) in the photograph of Figure~3  corresponds
to Figure~2. It consists of two perspex sheets with the profile of $z=\frac{1}{\sqrt{y}}$, which are the sides of the model; they are glued to a perspex square base of 21cm of side, and, along the curved profiles, to a flexible transparent plastic sheet.
We have used perspex sheets of 2mm of thickness and the profiles have been made by laser cutting in a plastics workshop. The two transversal pins can be toothpicks or plastic wires.

When our model is dipped into a soap solution, a soap film is formed between the two pins. Incidentally, if soap films between the pins and the lateral sides are also established, they should be removed by piercing with a stick.
We have drawn on the base of our model, the cycloid that rolls over the $x$-axis and passing through the base points of the pins.
We can see in the photograph of Figure~4, that
the horizontal section at $z=0$ of the soap film fits very well to the drawn arc of cycloid.{\footnotemark[12]}

\begin{figure}
\begin{center}
\includegraphics[width=26pc]{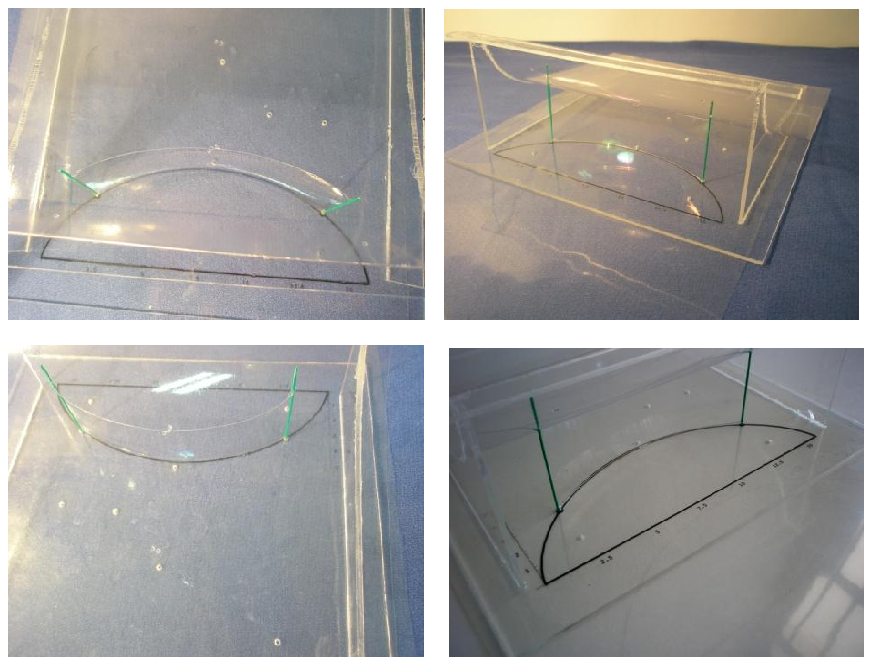}
\caption{Photographs of the soap film formed between surfaces $z=0$, $z=\frac{1}{\sqrt{y}}$, and the vertical pins over points $Q_1= (2.3,2.3)$ and $Q_2=(13.6,3.3)$ in cm. Its intersection with $z=0$ fits very well the arc of cycloid
passing through these two points and with base line the $x$-axis. This cycloid is
generated by a circle of radius $R=2.54$ cm rolling on the $x$-axis.}
\end{center} \label{Figure4}
\end{figure}

In our experiment, to minimize the gravity effects, we have assumed that the perspex base of our model has been placed horizontally. It can be instructive to see how the soap film is deformed when the base of the model is placed vertically.

\section{The solution to a generalized Steiner problem using soap films.}

The classical Steiner problem consists of finding a point $Q$ in the plane such that minimizes the sum of its distance to three given points $Q_1, Q_2, Q_3$,  that is,
\begin{equation}\label{Steiner}
    d(Q) = d (Q, Q_1) + d (Q, Q_2) +d (Q, Q_3)
\end{equation}
is minimized, where $d(Q, Q_i)$ is the Euclidean distance between $Q$ and $Q_i$, $(i=1,2,3)$.{\footnotemark[4]}
We assume that each angle of the triangle $Q_1Q_2Q_3$ is less than $120^{\circ}$.

This problem, that has many industrial and commercial applications, has an experimental solution using soap films.{\footnotemark[13]}
To see this, consider the soap film limited by two parallel plates placed horizontally and three vertical pins.
The soap film is composed of three vertical planes, which meet in a vertical edge over a point $Q$ at angles of $120^{\circ}$ to each other, according to the second Plateau law.{\footnotemark[8]}
 Because
the area minimization property of this soap film is equivalent to length minimization property of the base curve, this soap film gives the solution to the Steiner problem.

Consider now the Steiner problem for the cycloidal metric,  so that we have to minimize the sum of the length of three  cycloidal arcs that meet at certain point $Q$.
  The solution to this problem can be obtained experimentally by using soap films. For that, we only have to add to the perspex model described in Section~\ref{our model} a third vertical pin. Then, the base curve of the soap film that sets for this model, is the solution to this Steiner problem.

\begin{figure}
\begin{center}
\includegraphics[width=26pc]{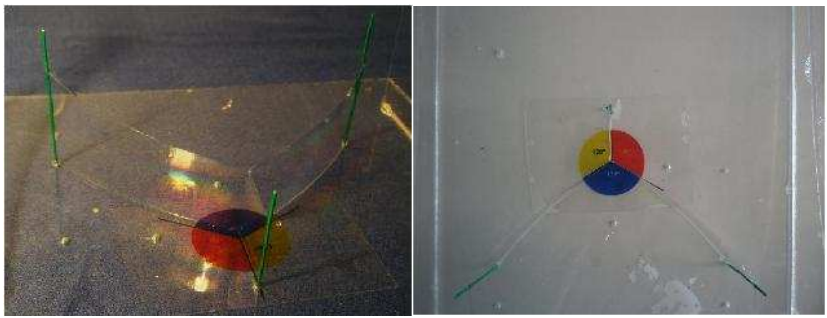}
\caption{Photograph of soap film solution to the Steiner problem for the cycloidal metric. Note that the three film components meet at angles of $120^{\circ}$ to each other.}
\end{center} \label{Figure5}
\end{figure}

 Figure~5 shows this experimental solution. Observe that the film components meet at angles of $120^{\circ}$ as in the classical Steiner problem.

\section{Solution to the problem of quickest approach with soap films.}
  The solution of Jakob Bernoulli to the Brachistochrone problem was published, with the other received solutions, in 1697 in Acta Eruditorum  under the title ``Solution of problems of my brother together with the proposition of other in turn".

In this article he proposed two new variational problems. The first is known as the problem of quickest approach, and the second is related with the isoperimetric problem.{\footnotemark[14]}

Jakob enunciated the first problem in these terms (see Figure 6):

\emph{On which of the infinitely many cycloids passing through $O$, with the same base $OA$, can a heavy body fall from $O$ to a given vertical line $AB$ in the shortest time?}

\begin{figure}
\begin{center}
\includegraphics[width=10pc]{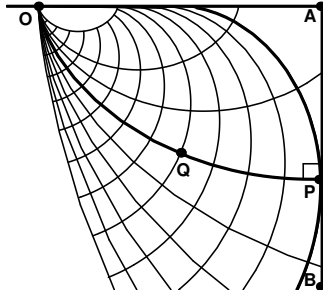}
\caption{The family of cycloids passing through $O$ and with the same base $OA$. The cycloid along which a heavy body falls from $O$ to $AB$  in the shortest time is precisely the arc $OQP$ that meets $AB$ orthogonally. The transversal curves are the synchronies, that is, the curves of the simultaneous positions of heavy particles which are released at $0$ along the cycloids at the same time.}
\end{center} \label{Figure6}
\end{figure}

His brother Johann solved this problem quickly, showing that the cycloid of quickest approach is the cycloid that meets orthogonally with the vertical straight line $AB$ (see Figure~6). He also solved the problem for the general case in which $AB$ is not a straight line. The solution is, also in this case, the cycloid orthogonal to $AB$.{\footnotemark[15]}

\begin{figure}
\begin{center}
\includegraphics[width=20pc]{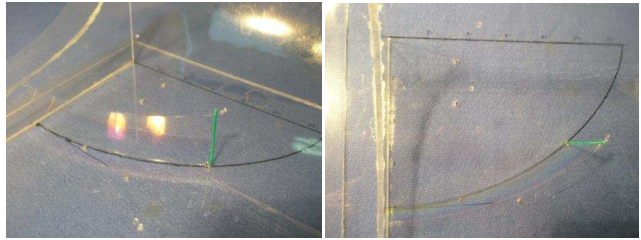}
\caption{Photographs of soap film solution to the problem of quickest approach. The pin is over the point $(4, 5.5)$ in cm. The soap film meets the vertical surface orthogonally (1st Plateau law), and its horizontal section at $z=0$ is an arc of cycloid  with $R=4$ cm. }
\end{center} \label{Figure7}
\end{figure}

In Figure~7 we can see the solution to this problem using soap films. We have used the model described in Section \ref{our model} consisting of two perspex plates corresponding to the surfaces $z=0$ and $z=\frac{1}{\sqrt{y}}$, and we have consider again, as a good approximation, that soap films between vertical pins or planes are vertical surfaces.
Translated to this model, the problem is to find the surface of minimal area that connects a given vertical pin with a given vertical surface.
The soap film solution corresponds approximately to a vertical surface with base a cycloidal arc that meets the given vertical surface at an angle of $90^{\circ}$. The configuration for other angle is not stable, as expressed by Plateau first law.{\footnotemark[8]}
The required arc of cycloid (see Figures~6 and~7) has a generating circle whose perimeter is twice the distance $OA$. Note that  $AP=2R$, $OA=\pi R$.

\section{The catenary and the Poincar\'e half-plane.}

\subsection{The catenary.}

\begin{figure}
\begin{center}
\includegraphics[width=22pc]{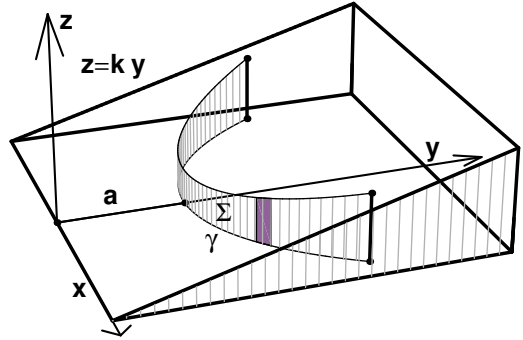}
\caption{Vertical surface $\Sigma$ of minimal area bounded by $z=0$, $z=ky$, and two vertical pins. Its horizontal section at $z=0$ is a catenary $\gamma$.}
\end{center} \label{Figure8}
\end{figure}

In 1691 Jakob Bernoulli proposed as a challenge to find the curve that assumes a freely hanging chain  when held down by its ends.
His brother Johann, Christiaan Huygens, and Gottfried Leibniz found the curve solution, since then known as the catenary.

As a variational problem, the catenary has to minimize the gravitational potential of the chain, that is:
\begin{equation}\label{potential}
   V(\gamma) = \int_{\gamma}gydm = g\rho\int_{\gamma}y ds = g\rho\int_{\gamma}y \sqrt{1+y'^2} dx
\end{equation}
where $gydm$ is the potential energy in an uniform gravitational field $g$ of an infinitesimal element $ds$ with mass $dm$ and density $\rho$, which is situated at altitude $y$ with respect to the point of zero potential energy level of reference.

The explicit equation for a chain of length $L$, two suspension points at the same altitude and with separation $H$,  can be written, for an adequate choice of the axis, as:
\begin{equation}\label{catenary}
    y= a\cosh \frac{x}{a}
\end{equation}
where $a$ depends on $H$ and $L$.{\footnotemark[16]}

On the other hand, the area of the vertical surface $\Sigma$, raised over a curve $\gamma$ (see Figure~8) is given by
\begin{equation}\label{area}
    A(\gamma) = \int_{\gamma}z ds = k \int_{\gamma}y \sqrt{1+y'^2} dx
\end{equation}

Since Eqs.~(\ref{potential}) and~(\ref{area}) only differ in a constant factor, the curve $\gamma$ that minimizes $V(\gamma)$ is the same that minimizes $A(\gamma)$, namely the catenary.

\begin{figure}
\begin{center}
\includegraphics[width=22pc]{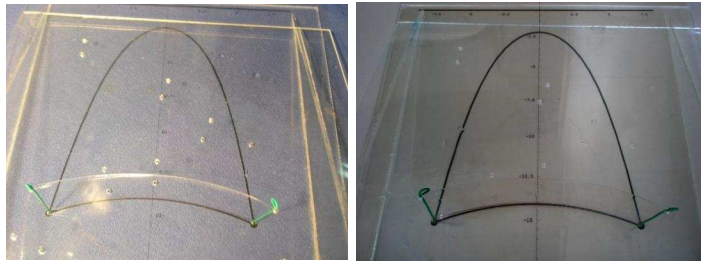}
\caption{Photographs of the soap film formed between $z=0$, $z= ky$, and the vertical pins over the points $(-5.6,-15.2)$ and $(5.6,-15.2)$ in cm. Its intersection with $z=0$ fits very  well the catenary $y= a\cosh (\frac{x}{a})$
with  $a=2.13$ cm. This is the stable solution (minimum energy). We have represented also the unstable solution (maximum energy) corresponding to $a=14.1$ cm.}
\end{center} \label{Figure9}
\end{figure}

If in our perspex model of Section 2 we replace the top curved surface $z= \frac{1}{\sqrt{y}}$ by the plane $z= ky$ with slope $k=\tan 15^{\circ}$ (see Figure 3 (b)), the soap film formed between two vertical pins is approximately a vertical surface whose base curve is a catenary (see Figures~8 and~9).{\footnotemark[12]}
 There are two solutions corresponding to a minimum and a maximum of elastic energy, so the first solution is stable while the second is unstable.{\footnotemark[17]}

 In our case, for the vertical pins over the points $(-5.6,-15.2)$ and $(5.6,-15.2)$ in cm, the two solutions calculated from Eq.~\ref{catenary} correspond to $a= 2.13$ cm and $a=14.1$cm.
The photographs of Figure~9 show that the experimental soap film fits very well the catenary drawn on the bottom plate according to the above equation for the stable case.

David Lovett also studied the catenary with the help of soap films by showing that the soap film formed between two pins across two non-parallel plates forming a wedge is part of a catenoidal surface.{\footnotemark[8]}{$\footnotesize{^, }$}{\footnotemark[18]}

\subsection{The Poincar\'e half-plane.}
After demonstrating that the cycloid is the curve that minimizes $\int_{\gamma}\frac{ds}{\sqrt{y}}$, Johann Bernoulli showed that the curve $\gamma$ that minimizes $\int_{\gamma}\frac{ds}{y}$ is an arc of circle. This means that, considering the Poincar\'e half-plane, that is, the half-plane $H^+ = \{ (x,y), y>0\}$ with the metric $\frac{ds}{y}=\frac{\sqrt{dx^2+dy^2}}{y}$,
the curve of minimal longitude joining two points $Q_1$ and $Q_2$ of $H^+$ that are not in a vertical line, is the only semicircle passing through $Q_1$ and $Q_2$ and with center a point $C$ on the axis $y=0$ (see Figure~10).{\footnotemark[19]}

\begin{figure}
\begin{center}
\includegraphics[width=20pc]{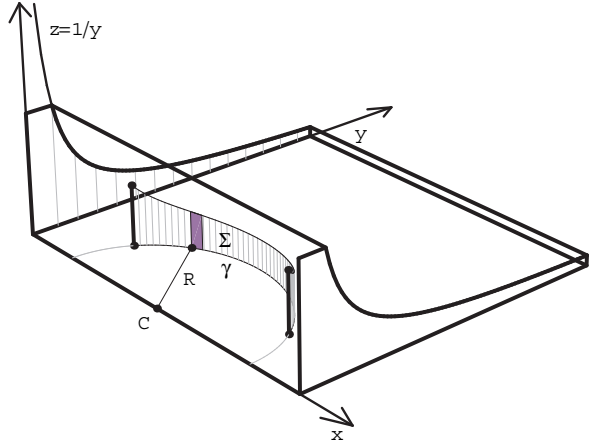}
\caption{Vertical surface $\Sigma$ of minimal area bounded by $z=0$, $z=\frac{1}{y}$, and two vertical pins. Its horizontal section $\gamma$ is an arc of circle with center in the $x$-axis.}
\end{center} \label{Figure10}
\end{figure}

Again, we can get the above solution using soap films.
For that, we have to replace the top surface of our perspex model with a surface of equation $z=\frac{1}{y}$ (see Figure~3(c)), so that the soap film that is formed between two vertical pins is approximately a vertical surface with base curve a semicircle (see Figures~10 and~11). This follows from the fact that the area of a vertical surface over $\gamma$ is given by:
\begin{equation}\label{semi}
    A(\gamma) = \int_{\gamma}\frac{ds}{y}= \int_{\gamma}\frac{\sqrt{1+y'^2}}{y}dx
\end{equation}
and therefore minimizing the area of $\Sigma$ is the same as minimizing the length of $\gamma$ in the metric $\frac{ds}{y}$.

\begin{figure}
\begin{center}
\includegraphics[width=20pc]{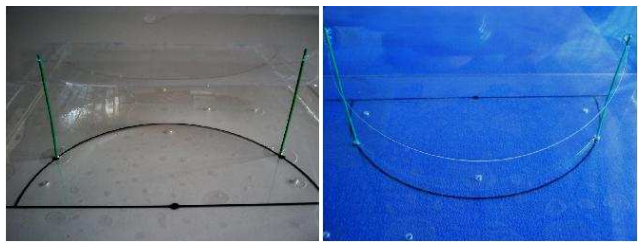}
\caption{Photographs of the soap film formed between $z=0$, $z= \frac{1}{y}$, and two vertical pins over the points $(-5.6,3.3)$ and $(5.6,3.3)$ in cm. Its intersection with $z=0$ fits very  well an arc of circle centered in the $x$-axis with radius $R=6.5$ cm.}
\end{center} \label{Figure11}
\end{figure}
We can see in Figure~11 that the experimental soap film fits very well the semicircle drawn in the bottom plate.

The soap film in the  Poincar\'e half-plane model is specially interesting for visualizing geometric properties of the cosmological model of the universe associated to a matter density less than the critical, because in this case the space has negative constant curvature so that the  Poincar\'e half-plane is a bidimensional model for this space.{\footnotemark[20]}

Here, as in the case of the cycloid, we can consider the soap film solution in terms of the Fermat principle for geometric optics, and consider the arcs of catenary and the arcs of circle as the path of the light rays in a medium with refractive index proportional to $y$ and $\frac{1}{y}$ respectively. We can also study soap films formed by three or more vertical pins, as experimental solutions of the associated Steiner problems for the metrics $yds$ and $\frac{ds}{y}$. The last case is of special interest because it gives an experimental solution to the Steiner problem in the Poincar\'e half-plane space.

Finally, it is interesting to observe that the cycloids and the arcs of circle can also be considered catenaries for gravitational potentials proportional to $\frac{1}{\sqrt{y}}$ and $\frac{1}{y}$, respectively.

\section{Summary}

We have seen how to design experiments with soap films formed between two perspex surfaces to study the solution to several variational problems that appear in mechanics, optics, elasticity, and geometry.

We have studied the brachistochrone problem, the Steiner problem, the problem of quickest approach, the catenary problem and the Poincar\'e half-plane problem. All are very important from a pedagogical point of view.

In our perspex models we have taken one of the surfaces as a plane ($z=0$) and for the other have considered three possibilities: $z= \frac{1}{\sqrt{y}}$, $z=k y$, and $z= \frac{1}{y}$, such that the soap films formed give the solution to the brachistochrone problem, the catenary problem, and the Poincar\'e half-plane problem, respectively. To prove this we have seen that the problem of minimizing certain mechanical magnitudes is equivalent to minimizing the area of a surface connecting two pins between two appropriate surfaces.

The above problems can also be considered as the optical problems of finding the path that follows a light ray, moving in a medium whose refractive index is proportional to $1/\sqrt{y}$, $y$, or $1/y$, respectively. In the context of geometry these problems consist in to calculate the geodesics for the metrics $ds/\sqrt{y}$, $yds$, or $ds/y$, respectively, where $ds=\sqrt{dx^2+dy^2}$ is the Euclidean metric.
Finally, the curves solution to the above problems (that is, arcs of cycloid, catenary and circle), can all be considered catenaries by taking gravitational potentials proportional to $\frac{1}{\sqrt{y}}$, $y$, and $\frac{1}{y}$, respectively.

These considerations illustrate in a simple way, the beautiful connection between dynamics, statics, optics, elasticity, and geometry.

Moreover, we have pointed out the historical origin of these problems, with the challenges of the brothers Jakob and Johann Bernoulli.

\section*{Acknowledgments}
This work was partially supported by the Ministerio de Educacion y Ciencia
grants TEC2007-60996/MIC (C. Criado) and MTM2007-60016 (N. Alamo),
 and by the Junta de Andalucia grant FQM-213 (N. Alamo).

\section*{Appendix A: Suggested problem}
Consider the following two surfaces: one is the horizontal plane $z=0$, and the other is defined by $z=\sqrt{y}$, $y>0$.
Consider also two vertical segments  connecting
these surfaces.
\begin{itemize}
\item[(a)] Prove that the vertical surface of minimal area bounded by these two segments and the two surfaces intersects the plane $z=0$ in an arc of parabola. For which positions of the segments there is not any solution? (Hint: The Euler-Lagrange equation gives two solutions but only one corresponds to a minimum.)
\item[(b)] Find the refraction index for which the light rays in $H^+=\{ (x,y); y>0\}$ describe parabolas.
\item[(c)] Find the gravitational potential for which the catenary is an arc of  parabola.
\item[(d)] What is the metric in $H^+$ for which the parabolas are the geodesics?
\end{itemize}

\vskip.5cm

\flushleft

\item[{$\footnotesize{^{a)}}$}]  {Electronic mail:
c\_criado@uma.es}

\item[{$\footnotesize{^{b)}}$}]  {Electronic mail:
alamo@uma.es}

\item[{\footnotemark[1]}]
Johann Bernoulli challenged the mathematical world with the brachistochrone problem, published in Acta Eruditorum in June, 1696, see D. J. Struick (ed), \emph{A source book in Mathematics}, (Princeton University Press, Princeton, New Jersey, 1986), pp. 391-392.

\item[{\footnotemark[2]}]
 To obtain this solution, he used the Fermat
Principle and considered that the curve of shortest descent time
must be the same that the curve described by a light ray whose
velocity is proportional to the square root of the altitude, that
is, the velocity that a heavy body acquires in falling.
This solution was published in Acta Eruditorum in May, 1697, pp. 206-211.
In the same issue of that journal, appeared also the solutions to the brachistochrone problem given by
his brother Jakob Bernoulli, Leibniz, L'H\^opital, Tschirnhaus, and  Newton. All except  L'H\^opital find the cycloid as the solution. The method used by Jakob Bernoulli contained the roots of the calculus of variations (see reference in Note 1).

\item[{\footnotemark[3]}]
 J. A. F. Plateau, \emph{Statique exp\'erimentale et th\'eorique des liquides soumis aux seules forces mol\'eculaires}, 2 Vols. (Gauthier-Villars, Paris, 1873).

\item[{\footnotemark[4]}]
R. Courant and H. Robbins, \emph{What is Mathematics?}  (Oxford University Press, London, 1941)

\item[{\footnotemark[5]}]
C. V. Boys, \emph{Soap-Bubbles, their colours and the forces that mould them} (Dover, New York, 1959).

\item[{\footnotemark[6]}]
S. Hildebrant and A. Tromba, \emph{Mathematics and Optimal forms} (Scientific American Books, New York, 1985).

\item[{\footnotemark[7]}]
C. Isenberg, \emph{The Science of Soap Films and Soap Bubbles} (Dover, New York, 1992).

\item[{\footnotemark[8]}]
D. Lovett, \emph{Demonstrating Science with Soap
Films} (IOP-publishing, Bristol, 1994).

\item[{\footnotemark[9]}]
A sufficient condition for a curve $\gamma$ to make the integral $\int_{\gamma}f(y,y',x) dx$ extremal is that it satisfies the second order differential equation $\frac{\partial f}{\partial y}- \frac{d}{dx}\frac{\partial f}{\partial y'}=0$, known as the Euler-Lagrange equation (see H. Goldstein, \emph{Classical Mechanics} (Addison-Wesley, Reading, Massachusetts, 1980), p.45.
For the case in which $f$ does not depends on $x$, the Euler-Lagrange equation reduces to
$y'\frac{\partial f}{\partial y'}- f = \textrm{constant}$. In our case, $f(y,y') = \frac{\sqrt{1+y'^2}}{\sqrt{y}}$ and then we get the differential equation $y' = \sqrt{\frac{k-y}{y}}$, where $k$ is a constant. It is easy to see that its  solution is the cycloid.

\item[{\footnotemark[10]}]
R. Osserman, \emph{A Survey on Minimal Surfaces} (Dover Publications, New York, 1986).

\item[{\footnotemark[11]}]
According to the first Plateau law (see, for example, Ref. 8, pp. 8-9), the film have to meet both surfaces $z=0$ and $z = \frac{1}{\sqrt{y}}$ at an angle of $90^{\circ}$. Therefore, the
approximation of taking a vertical surface $\Sigma$ is better as $y$ is larger, because then the slope of the tangent to $z=1/\sqrt{y}$ is smaller.

\item[{\footnotemark[12]}]
A  rough estimation of the relative error when we take a vertical element of surface $\Delta A$ instead of the element of surface for the soap film can be calculated by
\begin{equation}
  \epsilon_r = \frac{\Delta A' -\Delta A}{\Delta A} = \frac{(\alpha - \tan \alpha) \Delta s}{\tan \alpha\  \Delta s}
\end{equation}
where  $\Delta A'$ is the element of surface for a cylindrical surface that meets orthogonaly the two plates,  $\Delta s$ is the line element along the curve $\gamma$, and $\tan \alpha = \frac{d z}{d y}= z'$.

Neglecting the terms of order greater than 4 in the expansion of $\alpha - \tan \alpha$ as function of the derivative $z'$, we get $\epsilon_r \simeq -z'^2/3$ which for $z=1/\sqrt{y}$ leads to $\epsilon_r \simeq -y^{-3}/12$. Therefore, for $y>2$ we have an error less than $1\%$.

Similar calculations can be made for $z=ky$ and $z=1/y$ for the cases of the catenary and the Poincar\'e half-plane, respectively.

\item[{\footnotemark[13]}]
C. Isenberg, ``Problem solving with soap films," Phys. Educ. \textbf{10}, 452-456 (1975).

\item[{\footnotemark[14]}]
See reference in Note 1, p. 399

\item[{\footnotemark[15]}]
Johann Bernoulli could easily solve this problem because of his early study on the synchronies, which are the orthogonal trajectories to the cycloids passing through $O$ (see Figure~6). From a mechanical point of view, these curves are formed by simultaneous positions of heavy particles which are released at $O$ along the cycloids at the same time. On the other hand, from an optical point of view, the synchronies correspond to the simultaneous positions of the light pulses emitted from $O$ at the same instant and that propagate along a medium with refractive index proportional to $1/\sqrt{y}$.
Therefore the synchronies are the wavefronts.
From the fact that wavefronts and light rays are perpendicular, Johann Bernoulli concluded that the synchronies are the orthogonal trajectories to the cycloids. Thus the point of quickest approach $P$ must be the point where one of the synchronies is tangent to the line $AB$ (see Figure~6). For more details on this story see S. B. Engelsman, \emph{Families of Curves and the Origins of Partial Differentiation} (North-Holland, Amsterdam, 1984), pp. 31-37.

\item[{\footnotemark[16]}]
The parameter $a$ is determined by $\sqrt{a^2 +(\frac{L}{2})^2} = a\cosh \frac{H}{2a}$. The suspension points are symmetric with respect to the $y$-axis and their distance to the $x$-axis is $b=\sqrt{a^2 +(\frac{L}{2})^2}$. For the proof of these facts see
J. L. Troutman, \emph{Variational Calculus and Optimal Control} (Springer Verlag, New York, 1996),  pp. 78-80.

\item[{\footnotemark[17]}]
When the ratio between pin separation and distance from the pins to the $x$-axis, $\frac{H}{b}$, is greater than $1.335$  there is not any parameter $a$ verifying $b=a\cosh \frac{H}{2a}$ (see Note~16), while for $\frac{H}{b}$ less than $1.335$  there are two  values of parameter $a$.

\item[{\footnotemark[18]}]
 The catenoid is the surface of revolution generated by the catenary, and, as Euler showed in 1744, is the only minimal surface of revolution (see Ref.~6). The reason is that the area of a surface of revolution generated by a curve $\gamma$ of equation $y=y(x)$ is given by $A(\gamma) = 2\pi\int_{\gamma}y ds = 2\pi\int_{\gamma}y \sqrt{1+ y'^2}dx $, and again coincides, except for a constant factor, with Eqs. (\ref{potential}) and (\ref{area}). The Euler-Lagrange equation (see Note~9), reduces in this case to
$\frac{y}{\sqrt{1+y'^2}} = \frac{1}{c_1}$, and  its integration gives the catenary of equation $y= c_1\cosh \frac{x+c_2}{c_1}$.

\item[{\footnotemark[19]}]
The Euler-Lagrange equation (see Note~9), reduces in this case to
$\frac{1}{y \sqrt{1+y'^2}} = c_1$, and  its integration gives the semi-circle of equation $y= c_1\sin t , x= c_2+c_1\cos t$, for $ 0< t < \pi$.

The Poincar\'e half-plane is one of the simplest models of non-euclidean spaces. The geodesics of this space are the semicircles with center a point $C$ on the axis $y=0$ together with the vertical straight lines. If we think of these geodesics as the ``straight lines" of this space, the associated geometry does not verify the $5^{th}$ Euclidean postulate (the parallel postulate), because given a ``straight line" and a point not on that line, there are infinitelly many lines through the point that do not cut (i.e. parallels to) the given straight line. Therefore, only a small step, considering semicircles as ``straight lines" separated Johann Bernoulli to discover non Euclidean geometry 130 years before Bolyay and Lobatchesky did.

\item[{\footnotemark[20]}]
 W. Rindler,  \emph{Relativity: Special, General and
 Cosmological},
(Oxford University Press, New York, 2001), p.~364.

\end{document}